\documentclass[runningheads]{llncs}

\usepackage{comment}
\usepackage{environ}
\NewEnviron{hide}{\comment{\wrap}}
\NewEnviron{extra}{\comment{\wrap}}

\newif\ifshow 
\ifshow
  \includecomment{extra} 
  \excludecomment{hide} 
\else
  \excludecomment{extra} 
  \includecomment{hide} 
\fi

\usepackage{graphicx}
\usepackage{hyperref}
\usepackage{textcmds}
\usepackage{amsmath}

\usepackage{amsthm}
\newtheorem{observation}{Observation}

\usepackage{breakcites}
\usepackage{algorithm}
\usepackage{algorithmic}
\usepackage{caption}
 \makeatletter
\newenvironment{sketchproof}[1][]
{%
 \def\firstargument{#1}
 \def\noargumentspecified{}
 \ifx\firstargument\noargumentspecified
  \begin{proof}[Proof]
 \else
  \begin{proof}[Proof of \autoref{#1}]
  \label{proof:#1}
 \fi
 {}
} %
{\end{proof} }

\begin{document}
\title{Reactive PLS for Distributed Decision}
%
%
\author{Jiaqi Chen\inst{1} \and
Shlomi Dolev\inst{2}\thanks{Shlomi Dolev's work is supported by the Lynne and William Frankel Center for Computer Science, the Rita Altura Trust Chair in Computer Science, and is partially supported by a grant from the Ministry of Science and Technology, Israel \& the Japan Science and Technology Agency (JST), and the German Research Funding (DFG, Grant\#8767581199).} \and
Shay Kutten\inst{1}\thanks{The research of Shay Kutten was supported in part by a grant from the Hiroshi Fujiwara Cyber Security Research Center at the Technion.}}
\authorrunning{J. Chen et al.}
%
\institute{Technion - Israel Institute of Technology, Haifa, Israel\\
\email{kutten@technion.ac.il}\\ \and
Ben-Gurion University of the Negev, 84105 Beer-Sheva, Israel\\
\email{dolev@cs.bgu.ac.il}}
\maketitle              
\begin{abstract}
We generalize the definition of Proof Labeling Schemes to reactive systems, that is, systems where the configuration is supposed to keep changing forever. As an example, we address the main classical test case of reactive tasks, namely, the task of token passing. Different RPLSs are given for the cases that the network is assumed to be a tree or an anonymous ring, or a general graph, and the sizes of RPLSs' labels are analyzed. We also address the question whether an RPLS exists. First, on the positive side, we show that there exists an RPLS for any distributed task for a family of graphs with unique identities. For the case of anonymous networks (even for the special case of rings), interestingly, it is known that no token passing algorithm is possible even if the number $n$ of nodes is known. Nevertheless, we show that an RPLS {\em is} possible. On the negative side, we show that if one drops the assumption that $n$ is known,  then the construction becomes impossible. 

\end{abstract}
\section{Introduction}
\label{sec:intro}
Proof Labeling Schemes \cite{korman2010proof} as well as most later variations of distributed ``local''  verification of global properties, were suggested for ``input/output'' distributed algorithms, where the (distributed) input is available at the nodes before an algorithm is executed, and the algorithm uses it to generate the final output and terminate. 
(Those tasks are also called ``static'' or ``one-shot''; they are``silent'' in terms of self-stabilization \cite{dolev1999memory}).
As discussed in \autoref{sub:related}, they were shown useful in the context of the development of algorithms, as well as in the context of distributed computability and complexity. 
This motivates the extension formalized in the current paper -- that of local checking in {\em reactive systems}, where the configuration is supposed to keep changing forever. We note that reactive systems are by far, the bulk of distributed systems.

In the current paper, we define and initiate the development of efficient Reactive Proof Labeling Schemes (RPLSs) tailored to specific tasks. As a test case, we address the main classical problem of self-stabilization -- token passing. Different RPLSs are given for the cases that the network is assumed to be a tree, or a ring, or a general graph, and the sizes of their labels are analyzed. 

 We also address the question whether an RPLS exists. First, on the positive side, we show that there exists an RPLS for any distributed task for a family of graphs with unique identities. For the case of anonymous networks (even for the special case of rings), interestingly, it is known that no token passing algorithm is possible even if the number $n$ of nodes is known \cite{dijkstra}. (An algorithm is known for the special case that $n$ is prime \cite{burns1989uniform}). Nevertheless, we show that an RPLS {\em is} possible. On the negative side, we show that if one drops the assumption that $n$ is known,  then the construction becomes impossible.

Recall that by coupling a non-self-stabilizing algorithm with a ``one-shot'' task, one could develop a self-stabilizing algorithm for that task. Similarly, coupling such an RPLS with a non-stabilizing token passing algorithm would ease the task of designing a self-stabilizing token passing algorithm. Another possible direct application is to allow the use of a randomized algorithm until stabilization when deterministic algorithms do not exist, and to cease the use of potentially expensive randomness when the RPLS does not indicate an illegal state \cite{dolev2010randomization}.

\vspace{-0.3cm}

\subsection{Background}
\label{sub:related}
The notion of local checking started in the context of self-stabilization, and later motivated further research.
Early self-stabilizing algorithms \cite{dijkstra,lamport,DIM89IEEEWSS,burns1989uniform,bib349:SSODSAORWA1993}  as if by magic, managed to {\em stabilize}
(see definitions in \cite{dolev2000self}) without explicitly declaring the detection of a faulty global state. In \cite{dijkstra}, the configuration (the collection of all nodes' state) is ``legal'' iff the state of exactly one node is ``holding a token''. A node holding a token takes a certain step to ``pass the token'' to the next node, whether the configuration is legal or not.
This seems a much more elegant design than addressing multiple cases. Looking at this very elegant algorithm, one is amazed how
extra tokens (if such exist) somehow disappear. Similarly, in \cite{DIM89IEEEWSS,bib349:SSODSAORWA1993}, a node repeatedly chooses to point at the neighbor who has the lowest value in the ``distance-from-the-root'' variable whether the current configuration is legal or not. This, again, elegantly avoids cluttering the algorithm with various cases and subcases. Surprisingly,
if the configuration is not already a Breadth First Search tree rooted at the assumed special node, then it converges to being such a tree.

Despite the elegance of the above approach, it also has the disadvantage of requiring the ingenuity to especially design an algorithm for different tasks from scratch.
On the other hand, the opposite approach, that of detecting an illegal state and then addressing it, also seems difficult at first.  As Dijkstra described  \cite{dijkstra}, the difficulty in self-stabilization arises from the fact that the configuration is not known to anybody. A node is not aware of the current state of far away nodes. It can know only about some past states that may have changed.
In \cite{chandy1985distributed, katz1993self} a way was suggested for overcoming this inherent difficulty in distributed systems. A ``snapshot'' algorithm collects a ``consistent'' (see  \cite{chandy1985distributed})
representation of the configuration to some given ``leader'' node, who can then check whether the configuration follows some given \emph{legality predicate}, e.g., ``exactly one node has a token'', and detect the case where it does not. (A self-stabilizing algorithm to elect such a leader
was suggested at the same time in \cite{afek1990memory}.)
This could simplify the task of designing self-stabilizing algorithms by practically automating the part of the task involving the detection of a faulty state. This allows the modular approach of designing a part of the algorithm especially for addressing faults. However, this kind of checking, requiring that a fault even at very remote nodes is detected, was named ``global checking'' \cite{awerbuch1994self} for a good reason. It carried a lot of overhead in time and communication
(even if the communication was reduced by tailoring the global checking to a specific task and even assuming that the checking process itself could not suffer faults but only the checked task may be faulty \cite{lin1992observing,kor2011tight}).

The notion of {\em local} checking and detecting a violation of the legality predicate on the configuration was thus suggested in \cite{afek1990memory}. The main new idea (the ``local detection paradigm'') was that if the configuration was {\em not} legal, then {\em at least one node} detects the violation using {\em only} its own state and the state of its neighbors. (If the configuration was legal, then {\em no node} detects a violation). Note that the locality was made possible by allowing the detection to be made by a single and arbitrary node, which is enough in the context of self-stabilization, since then correction measures (of various kinds \cite{afek1990memory,awerbuch1994self,awerbuch1993time,awerbuch1991distributed,awerbuch1991self,afek2002local,dolev1995superstabilizing,ghosh1996fault,kutten1997time})  can be taken. An example of a predicate that can be checked that way is equality -- in the election algorithm of \cite{afek1990memory}, every node $v$ checked repeatedly whether the unique identity of the node the $v$ considered to be the leader was indeed the same as the identity its neighbors considered to belong to the leader. Predicates that can be computed that way are {\em locally checkable}.

Some other properties, e.g., the acyclicity of a subgraph, required enhancing the output (and the legality predicate) by the addition of information at each node to ensure local checkability of the enhanced predicate. This was generalized by various other papers, at first, still in the context of self-stabilization. In \cite{awerbuch1991distributed,awerbuch1991self}, general functions were discussed. In \cite{dolev1999memory}, this allowed the definition of a stronger form of self-stabilization by reaching a state in which communication can be saved. In \cite{beauquier1998transient}, the generalization was to consider different faults at different distances in time or in graph distance, and thus invest more only in cases where more exact pinpointing of the faults was needed.

The first impact of local checkability outside of the realm of self-stabilization is due to \cite{naor1995can}. Indeed, they motivated the study of locally checkable predicates by the fact that local detection is useful in self-stabilization. Still, the subject of their paper was the not-necessarily-self-stabilizing {\em computation} of a language (viewing the output configuration as a word in a distributed language) for which a locally checkable predicate holds. They asked which of those could also be computed in $O(1)$ time. This started a long line of research, eventually characterizing exactly what (distributed time) complexity classes exist for languages whose predicate is locally checkable. See e.g., \cite{balliu2020much}.

A further development beyond self-stabilization was the notion of Proof Labeling Schemes (PLS) \cite{korman2010proof}. Those were suggested as a distributed counterpart of the classical foundation of computing notion of Nondeterministic Polynomial (NP) time. Recall that a language is in NP iff there exists a ``verifier'' such that for each word in the language, there exists a ``witness'' that can be checked ``easily'' by the verifier. The word is in the language if and only if the verifier accepts. In a PLS, the {\em distributed} ``witness'' is some label at each node (possibly different in different nodes). The verifier is distributed in the sense that in each node, it can check only variables and the label of the node, and the labels of the node's neighbors.\footnote{Under some other definitions, the predicate is computed on the whole state of the neighbor \cite{afek1997local,dolev1999memory,goos2016locally} or just variables on a specific port of an edge at a node and the neighbor at the other endpoint of that edge \cite{awerbuch1991distributed,awerbuch1994self,awerbuch1991self}.}
Here, ``easily'' checking means $O(1)$ distributed time, rather than the polynomial non-distributed time of the classical NP notion.

This led to an ongoing rich area of research by adapting notions, and hopefully also results of classical non-distributed theory into the distributed realm, see e.g., \cite{fraigniaud2013towards,feuilloley2016hierarchy,kol2018interactive,naor2020power}. 
In some cases, there are multiple interactions between a prover and the distributed verifier at each node. This is still very different from the subject of the current paper. For example, in \cite{kol2018interactive,naor2020power}, multiple interactions are useful even in the verification of one, non-changing configuration, while here, we deal with a configuration that is supposed to be changed in an ongoing manner. Moreover, in  \cite{feuilloley2016hierarchy,kol2018interactive,naor2020power}, an all-powerful and non-distributed prover is assumed, that can communicate with the nodes. This is very different from the current paper where the idea is to make the proof labeling ``long-lived'' by having it maintained by a {\em distributed} mechanism such that the configuration changed by a correct algorithm can still be verified as correct after the change. \emph{Static} checking in a network where edges may be inserted or removed was addressed in \cite{foerster2017local}. However, the labels there are never changed, so this can support only the case that both the graph before the change and the graph after the change obey the predicate. Additional complexity related results \cite{sarma2012distributed} were obtained by expanding the research into less local forms of checking.
 
The local detection paradigm was also applied to obtain results in distributed testing, hardness of approximation, Peer-to-Peer networks, and more \cite{even2017three,sarma2012distributed,onus2007linearization, dolev2010randomization}. Many generalizations were suggested, such as verifiers who are allowed to communicate to a distance larger than 1, or to a given number of nodes \cite{fraigniaud2013towards}. Another line of research generalized the verifier to output some general function rather than a binary one (of either accepting or rejecting) \cite{emek2011online}.

Unfortunately, verification schemes were designed for a single configuration, e.g. the output of ``one time'' distributed tasks.
Algorithms that updated the labels were used, e.g. after a system reset \cite{afek1990memory,awerbuch1994self} or even every step \cite{afek1990memory}
(to verify that a certain global condition (a cycle) did not occur in the new configuration). Still, there, the tasks addressed were not reactive (e.g. leader election \cite{afek1990memory}).

Checking is indeed often intermixed in the design of reactive systems, to identify alarming situations that may need a repair.
More directly than that, we are inspired by the local stabilizer \cite{afek2002local} -- a general fault recovery mechanism.
Essentially, the stabilizer of  \cite{afek2002local}  reduces the distributed network into a non-distributed one by having each node $v$ maintain copies of the local states of every other node in the system (however, the copy of the state of a remote node is from an earlier time in history since it takes time until $v$ can hear about it). Hence, each node can simulate nodes in the whole system. (The simulations for the actions of the node's neighbors are until the very last time unit, while the simulation for the actions of further away nodes are until earlier times). The simulations performed by neighboring nodes are compared each step so inconsistencies (caused by a fault) are detected one time unit after the fault. In Section \ref{sec:universal} we use
the detection part of \cite{afek2002local} to develop a reactive counterpart (\autoref{sec:universal}) of the non - reactive ``universal'' PLS \cite{korman2010proof} since it too sends all the information available and can detect an illegal configuration for any predicate. Similar to the above universal PLS, the universal RPLS of Section \ref{sec:universal}  can be viewed more as a proof of possibility rather than an efficient algorithm, because of the large communication and memory required for replicating every piece of information everywhere continuously. 

\paragraph{Other related notions}
Another notion that sounds similar to an RPLS -- \emph{local observer} was proposed in \cite{beauquier2005observing,beauquier2005observing-prob}. A local observer algorithm enables the detection of stabilization
(versus the detection of illegality addressed here) through one observer at a unique node, as opposed to requiring one observer at every node in \cite{lin1992observing}. An observer, similar to an RPLS, does not interfere with the observed algorithm. However, the observers of \cite{beauquier2005observing,beauquier2005observing-prob,lin1992observing} relied on
the extra assumption that the variables and messages of the observing algorithm itself (at least in one node \cite{beauquier2005observing}) cannot be corrupted (possibly analogous to relying on extra assumptions when consulting an oracle in complexity theory).\footnote{Another extra assumption used in \cite{lin1992observing} was that of \cite{burns1989uniform} (prime size).}
 Such an assumption is not used here. Note that all these observers were {\em not} ``local'' in the sense used here, of detection time and distance.

  \paragraph{Paper structure:} In Section \ref{sec:prelim} we present definitions, including that of RPLS. In Section \ref{sec:universal} we prove that there exists an RPLS for any task on graphs with unique identities (if one disregards its cost). In Section \ref{sec:token}, we describe the general structure of an RPLS for token passing, a structure that will serve us later for token passing in several different families of graphs.
  In Section \ref{sec:tree} we present an RPLS for token passing on trees. It is much more efficient than the RPLS for token passing over rings, presented in Section \ref{sec:ring}, we also discuss there the case that an RPLS exists and the case it does not exists for anonymous rings. (We only address deterministic algorithms). Section \ref{sec:general} presents an RPLS for token passing on general graph. Finally, Section \ref{sec:discussion} contains concluding remarks.  
\section{Preliminaries}
\label{sec:prelim}
\subsection{Model}
We represent a distributed system by a connected graph $G = (V, E)$ that belongs to a graph family $F$. The topology of $G$ does not change. Let $n=|V|$. For any vertex $v \in V$, $N(v)$ is the set of neighbors of $v$ and $deg(v)$ is the number of neighbors of $v$. Either the nodes may be identical or each node $v$ may have a unique identifier $id(v)$ (that cannot be changed by the environment, nor by an algorithm). For the sake of simple presentation only, we use the ``KT1 model''\cite{awerbuch1990trade} 
\begin{extra}
(knowing neighbor's $id$)
\end{extra}
for trees 
and general graphs.
\begin{hide}
Each node $v$ knows the identifier of the neighbor at the other endpoint of each of $v$'s edges.
\end{hide} 
We use the sense-of-direction model \cite{flocchini2003sense}
for anonymous rings.

The communication model is synchronous and time is divided into rounds. In each round, a node goes through three stages in the following order: sending messages to neighbors, receiving messages sent by neighbors in the same round, and local computation. In the computation stage, each node $v$ follows some distributed algorithm $Alg$; let $v$'s {\em algorithm state} include all $v$'s variables and constants used by $Alg$, as well as the received messages. (We use just ``state'' when no confusion arises). 
 \begin{hide} and 
$Alg$ uses those to decide on $v$'s next state and the messages it will send in the beginning of the next round. 
\end{hide}

Let $s$ be a mapping from the nodes to the set of algorithm states. The collection of nodes' states is called the (algorithm) configuration. Denote by $V_s(r)$ (or just $V$ when $s$ is clear from the context) the set of nodes together with their local states for round $r$. Specifically, unless stated otherwise, when any value at round $r$ is mentioned, the reference is to the end of the computation stage of round $r$. For example, for each variable $var$ in a local state, we may use the notation $var_r$ to denote the value of $var$ at (the end of) round $r$. 
A \emph{legality predicate} $f$ is defined on $G_s(r) = (V_s(r), E)$. The value of $f$ may change when the configuration is changed by the environment (see ``fault'' below).  $Alg$ may also change the configuration in the computation stage. 

Often, it is convenient to define the legality predicate only for some subset of the variables of $Alg$ 
(see \cite{boaz-output-stab}), called the \emph{interface variables} (for example, any token passing algorithm can be revised slightly to also maintain a variable that says whether the node has the token or not). This way, predicate $f$ (and the RPLS defined below) can be made independent of the specific algorithm used (e.g., the specific token passing algorithm used) as long as it follows the specification for the problem it solves, and the specification is defined (only) on the interface variables. 
Henceforward, 
when speaking of the (algorithm) state of a node $v$, we may be speaking only of the interface variables at $v$. Similarly, the configuration may include just the interface variable of every node $v$. 
\begin{hide}
(Of course, in the extreme case, the interface variables may include all the variables of $Alg$).
Finally, 
$Alg$ may be any algorithm manipulating the given interface variables to follow $f$. 
\end{hide}

\begin{definition}
A \textbf{fault} occurs when the environment (rather than $Alg$) changes the value of at least one variable at one or more nodes. This can be either a variable in the algorithm state or a variable of the RPLS. 
\end{definition}

We assume that the environment may choose one round $r$ (only) such that every fault occurs just before round $r$ starts (and after round $r-1$ ends, if such exists). This constraint on the faults guarantees that any message a node $v$ receives from neighbor $u$ is indeed the one that $u$ sends to $v$ at the beginning of the same round. The constraint can be relaxed using known techniques (see the discussion of atomicity in \cite{dolev1993self} and elsewhere, e.g., \cite{afek1997local}), but is used here for the ease of the exposition. 

The execution of a reactive task is, in general, infinite. However, the definitions are simplified by choosing some arbitrary round $r_0$ and calling it the initial round, and calling its configuration the initial configuration.  

\begin{definition}
The execution is \textbf{correct} until $r$ (excluding $r$) if $\forall r_0 \leq r_k < r$, no faults occur before the beginning of $r$ and $f_{r_{\large k}}=1$.
\end{definition}

\begin{definition}
The configuration of $Alg$ becomes \textbf{incorrect} in $r$ if the execution is correct until $r$ (excluding $r$) but some faults occur just \emph{before} $r$ and change the configuration such that $f_r=0$.
\end{definition}
\subsection{Definition of Reactive Proof Labeling Schemes}
\begin{definition}
\label{def:rpls}
An RPLS for a predicate $f$ on some given set of interface variables is a pair $\pi = (\mathcal{M},\mathcal{V})$ of two algorithms. The  {\em marker} $\mathcal{M}$ algorithm 
consists of a \textbf{node marker} $\mathcal{M}(v)$ in each node $v$ that maintains a variable $L(v)$ called \textbf{a label} of $v$ which is stored in $v$ but is not considered to be a part of the (algorithm) state of $v$ and is not manipulated by $Alg$. The marker
is distributed\footnote{A non-distributed marker may also be useful in some settings.} but its specification includes the specification of its initial values at $r_0$ as a function of the  values of the interface variables at round $r_0$. 

Similarly, 
the {\em verifier} algorithm $\mathcal{V}$ consists of a \textbf{node verifier} $\mathcal{V}(v)$ in each node $v$.  
After $Alg$ completes its computation for round $r$, the marker $\mathcal{M}(v)$ may change the value of $L(v)$. Then,  $L(v)$ is sent to $v$'s neighbors at the beginning of round $r+1$. In the computation stage of $r+1$, verifier $\mathcal{V}(v)$ is executed before $Alg$.  The verifier uses $v$'s $Alg$ state, as well as $L(v)$ and $L(u)$ received from each neighbor $u$, to output either $0$ or $1$.

If the interface variables are only a subset of the $Alg$ state then, before the execution of $\mathcal{M}(v)$ in each round $r$, $Alg$ updates the values of $v$'s interface variables; they are accessible for $\mathcal{M}(v)$ in $r$ and for $\mathcal{V}(v)$ at the beginning of  $r+1$.

An RPLS $\pi = (\mathcal{M},\mathcal{V})$ \emph{is correct} if the following two properties hold.\\\indent
$i$. For every $G \in F$, if the execution is correct until $r+1$ (excluding $r+1$)
and the labels' values were computed according to the specification of $\mathcal{M}$ at round $r_0$,
then $\forall r_0 < r_k \leq r+1$, $\mathcal{V}(v)$ outputs $1$ for all $v$.\\\indent
$ii$. For every $G \in F$, if the configuration becomes incorrect in $r$, then $ \exists v \in G$ such that $\mathcal{V}(v)$ outputs $0$ at the beginning of $r+1$.
\end{definition}

 The interface between the RPLS and the algorithm is shown in \autoref{alg:main}. 
Intuitively, for each round, the RPLS implies a PLS (the specification of the initial labels plus the verifier's action). The addition here is that $\mathcal{M}$ updates the PLS each round
 to accommodate the actions of $Alg$. 
 If $f$ does not hold even in $r_0$, then definition \ref{def:rpls} implies that this is detected already in the next round (if no further faults occurred) and further updates of the labels do not interest us here. Hence, let us now treat the case that $f$ holds for the initial state.

\begin{definition}
A \textbf{legal initial labeling} for a given RPLS 
\begin{hide}
(of a  given $f$ on given interface variables
manipulated by some algorithm $Alg$) 
\end{hide}
is
the assumed set of labels of $\mathcal{M}(v)$ in a given initial configuration of $Alg$ at round $r_0$, defined if $f$ holds for that configuration.
\end{definition}

\begin{definition}
The \textbf{size} of an RPLS $\pi = (\mathcal{M},\mathcal{V})$ is the maximum number of bits in the label that $\mathcal{M}(v)$ 
assigns over all the nodes $v$ in $G$ and all $G \in F$ and all the rounds.
 For a family $F$ and a legality predicate $f$, we say that the \textbf{proof size} of $F$ and $f$ is the smallest size of any RPLS for $f$ over $F$.
\end{definition}
Note that the size of a label does not depend on $Alg$ (as long as $f$ does not depend on $Alg$, e.g., when $f$ is defined over the interface variables only).

\begin{algorithm}
\caption{Implementation of an RPLS and $Alg$ at $v$ at round $r$.}
\label{alg:main}
\begin{algorithmic}[1]
\STATE Send $L_{r-1}(v)$ to neighbors
\STATE Receive $L_{r-1}(u)$ from each neighbor $u$
\STATE Call $\mathcal{V}(v)$ 
\STATE Call $Alg$ (* this also updates the values of the interface variables *)
\STATE Call $\mathcal{M}(v)$
\end{algorithmic}
\end{algorithm}

\begin{hide}
\autoref{alg:main} describes the general workflow of an RPLS and $Alg$. Different verifier $\mathcal{V}(v)$ and  marker $\mathcal{V}(v)$ may be defined for different predicates or graph family. 
\end{hide}

For the RPLSs given in the next sections of this paper, we show the following two theorems:
\begin{theorem} 
\label{thm:correct}
Given (1) the RPLS is executed with $Alg$ and with a legal initial labeling at time $r_0$,  and (2)
the execution is correct until $r+1$ (excluding $r+1$), then all the verifiers output 1 in $r+1$.
\end{theorem}

\begin{theorem}
\label{thm:incorrect}
If the configuration becomes incorrect in $r$, then at least one verifier outputs 0 in $r + 1$.
\end{theorem}

\begin{observation}
By the assumption that if faults occur then they occur only between two rounds, what every node $v$ sends to $v$'s neighbors at the beginning of a round is the same as what the neighbors receive at the beginning of the same round. In addition, for \autoref{thm:incorrect}, ``the configuration becomes incorrect in $r$'' implies that faults occur only before $r$ and not after $r$. This ensures that, for every node $v$, label $L(v)$ generated at the end of $r$ is indeed the same as what $v$ sends to $v$'s neighbors at the beginning of $r+1$ and the same as what the neighbors receive at the beginning of $r+1$.
\end{observation}
This observation is used in the proofs of \autoref{thm:correct} and \autoref{thm:incorrect} and it might not be mentioned again in the relevant proofs.

\section{Universal RPLS}
\label{sec:universal}
In this section, we present a universal RPLS using ideas from the distributed local monitoring algorithm (call it local monitor) presented in \cite{afek2002local}. It is universal in the sense that it is a reactive proof labeling scheme for any algorithm. The down side is a possibly huge cost in terms of storage and communication.\\
\textbf{Short description of the local monitor:}\\
In a synchronous system (same as the one presented in this paper), a partial snapshot $VI_p^t(l)$ for node $p$ contains the collection of states and inputs of all the nodes that are within $l$ hops away from $p$ at round $t-l$. Each node $p$ at round $t$ maintains a pyramid $\Delta_p^t = VI_p^t(0), VI_p^t(1),...,VI_p^t(d)$ where $d$ is the diameter of the system. In the beginning of every round $t$, every two neighboring nodes $p$ and $q$ exchange their pyramids $\Delta_p^t$ and $\Delta_q^t$, then the local monitor at each node $p$ executes a verifying routine --- the value of $\Delta_p^t$ and $\Delta_q^t$ for every neighbor $q$ are checked against several predefined constraints, if one of the constraints is violated then the local monitor outputs ``INCONSISTENCY DETECTED'' and terminates, otherwise an updating routine is called --- it updates the value of $\Delta_p^t$ which are sent to $p$'s neighbors in the beginning of the next round. Informally, the conditions $p$ checks means that each node $q$ according to $q's$ state at each time $t'$ as known to $p$ indeed took the steps the algorithm dictates for $q$'s state. The check is by examining the states of $q$ and its neighbors, as known to $p$, for time $t'+1$. 

A universal RPLS is now constructed using the same idea. The label for each $p$ is $L(p) = \Delta_p^t$. 
To create the legal initial labeling, informally, simulate the first $d$ steps of some possible (legal) execution of $Alg$. This defines some state of each node for each time in  $0,1,2, ... , d$. Now, construct $VI_p^t(l)$ to include the states and inputs of all the nodes that are within $l$ hops away from $p$ at round $t-l$ according to the simulated execution.\footnote{This may sound not unlike the claim that the world was created only a few thousand years ago, but created with a history built in, e.g., looks as if there were dinosaurs in time much earlier than a few thousand years ago.}  Let us note that the such simulations of a fictitious execution proved useful also in generating the legal initial labeling of other RPLSs in this paper.  

More formally, legal initial labeling is one that consists of a valid pyramid which corresponds to a legal execution (see Definition 3.4 of \cite{afek2002local}) for each node such that all the constraints of the verifying routine of the local monitor are satisfied for every node $p$. 
The interface variables include any $Alg$ variable that is used by the local monitor. The algorithms for $\mathcal{V}(p)$ (\autoref{universalverifier}) and $\mathcal{M}(p)$ (\autoref{universalmarker}) are specified below.
\begin{algorithm}
\caption{$\mathcal{V}(p)$ at node $p$ for round $t$}
\label{universalverifier}
\begin{algorithmic}[1]
\STATE Send $L(p) = \Delta_p^t$ to $p$'s neighbors
\STATE Call the verifying routine (line 03-09 of Figure 1 in \cite{afek2002local}) with $L(p)$ and $L(q)$ received from each neighbor $q$ as input
\STATE \textbf{IF} the verifying routine outputs ``INCONSISTENCY DETECTED'': output 0
\STATE \textbf{ELSE}
        \STATE $\mathcal{V}(p)$ output 1
\end{algorithmic}
\end{algorithm}

\begin{algorithm}
\caption{$\mathcal{M}(p)$ at node $p$ for round $t$}
\label{universalmarker}
\begin{algorithmic}[1]
\STATE Call the updating routine (line 10-12 of Figure 1 in \cite{afek2002local}) to update $L(p)$
\end{algorithmic}
\end{algorithm}

\begin{sketchproof}[thm:correct]
By Theorem 3.2 of \cite{afek2002local}, the updating routine produces a valid pyramid (label) for each node in each round (if the execution is correct). By the definition of ``valid pyramid'' (Definition 3.4 of \cite{afek2002local}), it is a pyramid of snapshots that corresponds to the legal (correct) execution. It is implied that the collection of the valid pyramids satisfies the constraints of the verifying routine. Then it is easy to see that, if the execution is correct until $r+1$ (excluding), valid pyramids (legal labels) are exchanged in $r+1$ and none of the constraints of the verifying routine is violated in $r+1$ and every verifier in $r+1$ outputs 1 (line 4-5 of \autoref{universalverifier}).
\end{sketchproof}

\begin{sketchproof}[thm:incorrect]
By Theorem 3.1 of \cite{afek2002local}, if the configuration becomes incorrect in $r$ and no messages are corrupted in $r$, then one of the nodes outputs ``INCONSISTENCY DETECTED''. By line 3 of \autoref{universalverifier}, at least one verifier outputs 0.
\end{sketchproof}

\begin{observation}
For any distributed algorithm on a graph family with unique node identities,
there exists an RPLS.
\end{observation}
\section{RPLS for Token Passing Algorithms}
\label{sec:token}
\begin{definition}
Given a graph $G=(V, E)$ that belongs to a graph family F, a \textbf{token predicate} $f_{token}$ is a predicate evaluated at each node on its state and the messages it has received which determines whether it \emph{holds} a token.
\end{definition}

\begin{definition}
A \textbf{Token Passing Algorithm (TPA)} is any distributed algorithm for which a token predicate is defined.\\
The legality predicate \textbf{\textit{f}} for token passing evaluates to $1$ if there is exactly one token-holder on $G$ and evaluates to $0$ otherwise.\\
We assume that (in an execution with no faults) if TPA causes $f_{token}$ at some node $v$ to cease to hold at some round $r$ then TPA causes $f_{token}$ to start to hold at some neighbor $u$ of $v$. 
We then say that 
{\em the token is
passed from $v$ to $u$ in $r$}.

\end{definition}

We assume that the token can be passed exactly once from the token-holder $v$ to exactly one of $v$'s neighbors in each round until $r+1$ (excluding $r+1$)
if the execution is correct until $r+1$ (excluding $r+1$). 
In each round $r$, at line 4 of \autoref{alg:main}, TPA sets the interface variable $s_r(v) = 1$ if $v$ holds the token in $r$ and $s_r(v)=0$ otherwise. We assume $\mathcal{M}(v)$ knows whether the token is passed from some $u \in N(v)$ to $v$ or from $v$ to $u$ in $r$ when such a movement happens in $r$ (implemented easily using the notion of interface variables). 
\section{RPLS for Token Passing on Trees $T$}
\label{sec:tree}
Let us start with a simple example of a token passing over a tree. The new RPLS is a ``link reversal'' algorithm. Such algorithms maintain and change virtual directions on the edges, thus creating a directed graph rooted at objects such as a token, see e.g. \cite{welch2011link,ginat1989tight, demmer1998arrow, defago2020communication}. 
In the RPLS, the violation of $f$ can be detected by some node not having an edge directed out but still not having the token, or a node having an edge directed out but having a token, or by a node having two such edges. When the token is passed from a node $v$ to its neighbor $u$, the marker algorithm updates the direction of edge $(u,v)$ so that the tree is still rooted at the token. The nice property is that this update can be performed in the same round as the one in which a token is passed and only at the two nodes that are anyhow involved in the token pass. 

\begin{hide}
For the sake of simplicity only, we assume unique node identifiers $id(v)$. It is easy to change the RPLS to accommodate anonymous nodes. We also assume that there are $n$ nodes in the system but $n$ is not known to the nodes.
In addition, the
\end{hide}
\begin{extra}
The
\end{extra}
 RPLS maintains the following variables for each node $v$:\\
\textbullet $weight(v)_u$ = $0$, $1$ or $2$, $\forall u \in N(v)$ --- Informally, each edge $(u,v)$ is directed from the endpoint with the ``higher''(modulo 3) weight to the endpoint with the ``lower'' (modulo 3) weight. See \autoref{lcr} and \autoref{direction} below.\\
\textbullet $L(v) = (id(v), {(id(u), weight(v)_u) | \forall u \in N(v)})$.

\begin{definition}
\label{lcr}
For $0$, $1$ and $2$, define: $0 \prec 1, 1 \prec 2, 2 \prec 0$.
\end{definition}

\begin{definition}
\label{direction}
We say that edge $(v, u)$ is an \textbf{incoming} edge for $v$ and an \textbf{outgoing} edge for $u$ in round $r$, denoted by \emph{$v \xleftarrow{r} u$} if the following holds: for the pair of neighbors $u$ and $v$, there is one pair $(id(x), weight_r(v)_x)$ in $L_r(v)$ and one pair $(id(y), weight_r(u)_y)$ in $L_r(u)$ such that: $id(u) = id(x) \land id(v) = id(y) \land weight_r(v)_x \prec weight_r(u)_y$.
\end{definition}

\noindent {\bf Specification of the legal initial labeling:} denote the token holder of $r_0$ by $v$.
\begin{hide}
Recall that node $v$ has $s_{r_0}(v) = 1$ while every node $u \neq v$ has $s_{r_0}(u) =0$).
\end{hide}
The nodes are labeled such that $v$ has only incoming edges in $r_0$ while every node $u \neq v$ has exactly 1 outgoing edge and every other edge of $u$ is incoming to $u$. Clearly, the verifiers output 1 in $r_0+1$ (See C1 and C2 of \autoref{alg:treeverifier}).

\begin{algorithm}[H]
\caption{$\mathcal{V}(v)$ in round $r$}
\label{alg:treeverifier}
\begin{algorithmic}[1]
\STATE \textbf{IF} any of the following conditions is violated, output 0:
\begin{ALC@g}
\STATE C1: if $s_{r-1}(v) = 1$, then $\forall u \in N(v), v \xleftarrow{r-1} u$
\STATE C2: if $s_{r-1}(v) = 0$, then $\exists u \in N(v): u \xleftarrow{r-1} v \land \forall w \in N(v) \land w \neq u: v \xleftarrow{r-1} w$
\end{ALC@g}
\STATE \textbf{ELSE}, output 1
\end{algorithmic}
\end{algorithm}

\begin{algorithm}{H}
\caption{$\mathcal{M}(v)$ in round $r$}
\label{alg:treemarker}
\begin{algorithmic}[1]
\STATE \textbf{IF} token is passed from $u \in N(v)$ to $v$ in $r$:
    \begin{ALC@g}
    \STATE $weight(v)_u \Leftarrow (weight_{r-1}(u)_v - 1) \bmod 3$
    \end{ALC@g}
\end{algorithmic}
\end{algorithm}

\begin{hide}
\subsection{Proof of Correctness}

\begin{sketchproof}[thm:correct]
Clearly, there exists a single token in $r_0$, all the verifiers output $1$ in round $r_0+1$. Assume for induction that the execution is correct until $r$ (excluding $r$) and all the verifiers output $1$ in round $r$.\\
\textbf{Case 1:} The token is not passed in round $r$. Clearly, all the labels remain unchanged by \autoref{alg:treemarker} and the correctness for round $r + 1$ follows from the correctness for $r$.\\
\textbf{Case 2:} The token is passed from a node $u$ to a neighbor $v \in N(u)$ at round $r$. The marker $\mathcal{M}(v)$ updates $weight_r(v)_u \Leftarrow (weight_{r-1}(u)_v - 1) \bmod 3$ while $weight(u)_v$ remains unchanged ($weight_r(u)_v \Leftarrow weight_{r-1}(u)_v$) by $\mathcal{M}(u)$. TPA updates $s_r(v) \Leftarrow 1$ and $s_r(u) \Leftarrow 0$. Any other node remains in the same state and its label does not change in $r$ compared to $r-1$. In round $r+1$, verifier $\mathcal{V}(v)$ ``sees'' $s_r(v) = 1$ and $\mathcal{V}(u)$ ``sees''  $s_r(u) = 0$. They both ``see'' $v \xleftarrow{r} u$ ($weight_r(v)_u \prec weight_r(u)_v$) and output $1$ since constraints C1 and C2 are satisfied. The verifier at any other node $w$ ($w \neq u$ or $v$) also outputs $1$ in $r+1$ since the relation (implied by their labels) between $w$ and $w$’s neighbors did not change compared to round $r$ and the induction assumption was that they output 1 at round $r$. The theorem follows.
\end{sketchproof}

\begin{sketchproof}[thm:incorrect]
Assume by contradiction that all the verifiers output $1$ in $r + 1$.\\
\textbf{Case 1:} No token exists in $r$.  By our assumption, constraint C2 of all the verifiers is satisfied by \autoref{alg:treeverifier}, which implies that every node has exactly one outgoing edge. Since any two neighboring nodes $u$ and $v$ see the same value in $L_{r-1}(v)$ and $L_{r-1}(u)$, if edge $(u, v)$ is outgoing for $u$ then it cannot be an outgoing edge for $v$. Since there are $n$ nodes in the system, this gives us $n$ outgoing edges in the system. However, there are at most $n-1$ edges on the tree. A contradiction.\\
\textbf{Case 2:} Multiple (more than 1) tokens exist in round $r$. Case 2.1: Assume that two token-holders $u$, $v$ are neighboring nodes. Since $u$ and $v$ see the same value of $L_r(u)$ and $L_r(v)$ in $r+1$, the orientation of edge $(u, v)$ must be either outgoing for $u$ or $v$ and this contradicts either C1 of $\mathcal{V}(u)$ or C1 of $\mathcal{V}(v)$. Case 2.2: Assume that there are no neighboring tokens. There exists a pair of token-holders $p_1$ and $p_k$ such that the path $p_1$-$p_2$-…-$p_k$ on $T$ contains no other token-holders. Since $p_1$ is a token holder in $r$ and TPA sets $s_r(p_1) \Leftarrow 1$ in $r$, then $\mathcal{V}(p_1)$ must determine that $p_1$ has only incoming edges in $r+1$ since C1 is satisfied by our assumption, which means $p_1 \xleftarrow{r} p_2$. Since $p_2$ is not a token-holder $\mathcal{M}(p_2)$ sets $s_r(p_2) \Leftarrow 0$. Since C2 is satisfied by our assumption, verifier $\mathcal{V}(p_2)$ must determine that $p_2$ has exactly one outgoing edge. Therefore, edge ($p_2$, $p_3$) must be incoming to $p_2$. By induction on the order of the nodes on this path, one concludes that edge ($p_{k-1}$, $p_k$) is an incoming edge for $p_{k-1}$ and an outgoing edge for $p_k$. However, $p_k$ being a token holder must have only incoming edges because of C1. A contradiction.
\end{sketchproof}
\end{hide}

\section{RPLS for Token Passing on Anonymous Rings $R$}
\label{sec:ring}
Consider a ring $R = (V, E)$  of $n > 2$ identical nodes (without unique identifiers). We assume that $n$ is known to every node. Each node has two edges, each connecting to one neighbor. 
\begin{hide}

All the nodes have a common sense of direction which means that each 
\end{hide}
\begin{extra}
Each 
\end{extra}
node can distinguish between its two neighbors -- a {\em successor} (the clockwise neighbor) and a {\em predecessor} (the counterclockwise neighbor). The label of each node $v$ is an integer in $[0, n^2-1]$.

\begin{definition}
For any pair of integers $x$, $y$ in $[0, n^2-1]$, define $x \prec y$ if $y = (x + 1) \bmod n^2$
\end{definition}

\noindent \textbf{Specification of the legal initial labeling:} For any $k=0$,$1$,..., $n-1$, let node $v_{k+1 \bmod n}$ be the successor of $v_k$, such that $s(v_{n-1}) = 1$. (The numbering is not known to the nodes). 
The legal initial labeling is: For every $0 \leq k \leq n-1$, let $L(v_k) = k$. This gives us $L(v_{n-1}) = (L(v_0)+ n - 1) \bmod n^2$ and $L(v_0) \prec L(v_1) \prec ... \prec L(v_{n-1})$. Clearly, $f_{r_0} = 1$ and indeed all the verifiers output 1 in $r_0+1$ if no fault occurs before $r_0+1$. See C1 and C2 of \autoref{alg:ringverifier}.

\begin{algorithm}
\caption{$\mathcal{V}(v)$ in round $r$}
\label{alg:ringverifier}
\begin{algorithmic}[1]
\STATE /*Denote the successor of $v$ by $w$*/
\STATE \textbf{IF} any of the following conditions is violated, output 0:
    \begin{ALC@g}
    \STATE C1: if $s_{r-1}(v) = 1$, then $L_{r-1}(v) = (L_{r-1}(w) + n - 1) \bmod n^2$ 
    \STATE C2: if $s_{r-1}(v) = 0$, then $L_{r-1}(v) \prec L_{r-1}(w)$
     \end{ALC@g}
\STATE \textbf{ELSE} output 1
\end{algorithmic}
\end{algorithm}

\begin{algorithm}
\caption{$\mathcal{M}(v)$ in round $r$}
\label{alg:ringmarker}
\begin{algorithmic}[1]
\STATE /*Denote the predecessor of $v$ by $u$*/
\STATE \textbf{IF} token is passed from $u \in N(v)$ to $v$ in $r$: /*token passed clockwise*/
    \begin{ALC@g}
    \STATE $L(v) \Leftarrow (L_{r-1}(u) + 1) \bmod n^2$
    \end{ALC@g}
\STATE \textbf{IF} token is passed from $v$ to $u \in N(v)$ in $r$: /*token passed counterclockwise*/
    \begin{ALC@g}
    \STATE $L(v) \Leftarrow (L_{r-1}(u) - (n - 1)) \bmod n^2$
    \end{ALC@g}
\end{algorithmic}
\end{algorithm}

\begin{hide}
\subsection{Proof of Correctness}

\begin{sketchproof}[thm:correct]
The verifiers output $1$ in round $r_0+1$. Assume for induction that the execution is correct until $r$ (excluding $r$) and all the verifiers output $1$ in round $r$. \\
\textbf{Case 1:} The token is not passed in round $r$. Clearly, no labels change in $r$, and the correctness for round $r + 1$ follows from the correctness for $r$.\\
\textbf{Case 2:} The token is passed clockwise in $r$. Denote by $u$ the token holder in $r-1$, and denote $u$’s successor by $v$. First, we prove that $\mathcal{V}(u)$ and $\mathcal{V}(v)$ output $1$ in round $r + 1$. Denote $v$’s successor by $w$. By The induction hypothesis, we know that C1 and C2 of \autoref{alg:ringverifier} are satisfied in $r$, which implies $L_{r-1}(v) = (L_{r-1}(u) - (n - 1)) \bmod n^2$ and $L_{r-1}(w) = (L_{r-1}(v) + 1) \bmod n^2$. In round $r$, since the token is passed from $u$ to $v$, we have $L_r(v) \Leftarrow (L_{r-1}(u) + 1) \bmod n^2$, $s_r(u) \Leftarrow 0$, $s_r(v) \Leftarrow 1$ while the labels of $u$ and $w$ remain unchanged. Therefore, we have $s_r(u) = 0 \land L_r(u) \prec L_r(v)$ and $s_r(v) = 1 \land L_r(v) = (L_r(w) + n - 1) \bmod n^2$, which satisfies C1 and C2 for $\mathcal{V}(u)$ and $\mathcal{V}(v)$ in $r+1$. Both verifiers output $1$. Second, we prove that $\forall$ node $p \neq u$ and $p \neq v$, $\mathcal{V}(p)$ outputs $1$ in round $r+1$. For any node $p \neq u$ and $p \neq v$, denote $p$’s successor by $q$. Note that TPA updates $s_{r-1}(p)=0$ because $p$ does not hold the token in $r-1$ (since $u$ is the only token holder in $r-1$). Since $\mathcal{V}(p)$ outputs $1$ in round $r$ by the assumption, we know that $s_{r-1}(p) = 0 \land L_{r-1}(p) \prec L_{r-1}(q)$ (C2). In round $r$, no changes are made to $s_r(p)$, $L_r(p)$. In addiction, $L_r(q)$ remains unchanged. Therefore, the correctness for $r+1$ follows from the correctness for $r$ and every $p$ outputs $1$ in round $r + 1$.\\
\textbf{Case 3:} the token is passed counterclockwise in $r$. The proof of \textbf{\textbf{Case 3}} is similar to that of \textbf{Case 2}.
\end{sketchproof}

\begin{sketchproof}[thm:incorrect]
Assume for contradiction that all the verifiers output $1$ in $r+1$. For $k=0$,$1$,..., $n-1$, let node $v_{k+1 \bmod n}$ be the successor of $v_k$.\\
\textbf{Case 1:} No token exists in round $r$. For every node $v_k$, $s_r(v_k) = 0 \land L_r(v_k) \prec L_r(v_{k+1 \bmod n})$ since C1 and C2 are satisfied by the assumption. We have
$ L_r(v_0)= (L_r(v_{n-1}) + 1) \bmod n^2
\Rightarrow L_r(v_0)= (L_r(v_{n-2}) + 2) \bmod n^2
\Rightarrow L_r(v_0)= (L_r(v_0)+n) \bmod n^2
\Rightarrow n=1,-1 \Rightarrow$ It is a contradiction to $n>2$. \\
\textbf{Case 2:} Multiple (more than 1) tokens exist in round $r$. Denote by $i \leq n$ the number of tokens in the system ($i = 2$, $3$,..., $n$), then we have $n-i$ nodes without tokens. Since all the verifiers output $1$ in round $r+1$ by the assumption, both C1 and C2 are satisfied. For any node $v_k$ and $v_k$’s successor $v_{k+1 \bmod n^2}$, we have the relation following from C1 and C2:
\begin{align*}
L(v_k) & = [L(v_{k+1 \bmod n^2}) + s(v_k) \cdot (n - 1) + (1-s(v_k))\cdot( - 1)] \bmod n^2\\
& = [L(v_{k+1 \bmod n^2}) + n \cdot s(v_k) - 1] \bmod n^2\\
& = [L(v_{k+2 \bmod n^2}) + n \cdot s(v_{k+1 \bmod n^2}) + n \cdot s(v_k) - 2] \bmod n^2 \\
& =...\\
& = [L(v_k) + n \cdot i - n] \bmod n^2\\
& = [L(v_k) + n \cdot (i - 1)] \bmod n^2\\
\end{align*}
It implies that $n \cdot (i-1)$ is a multiple of $n^2$, and thus $i-1$ is a multiple of $n$. Therefore, $i-1 \geq n$ or $i-1 \leq -n$. However, since $2 \leq i \leq n$ we have $0 < i-1 < n$. A contradiction.
\end{sketchproof}
\end{hide}

\subsection{Necessity of Assumptions for any RPLS for anonymous rings.}
We prove that if one removes the assumption that each node knows the size of the ring, no RPLS exists for token passing on anonymous rings.
\begin{lemma}
\label{lem:reduction1}
If there exists an RPLS for $f$ on anonymous rings, then there exists a PLS for $f$ on anonymous rings.
\end{lemma}

\begin{hide}
\begin{sketchproof}
Recall that the main difference of an RPLS from a PLS lies in the marker $M$ that may change the label to accommodate the actions of $Alg$. Assume that we have an RPLS for $f$ on anonymous rings, if we use an algorithm $A_{static}$ that never moves the token, then the RPLS works exactly like a PLS.
\end{sketchproof}
\end{hide}

\begin{lemma}
\label{lem:reduction2}
If there exists a (static) PLS for $f$ on anonymous rings, then there exists a (static) PLS which verifies whether the size of an anonymous ring is $n$.
\end{lemma}

\begin{hide}
\begin{sketchproof}
We assume that there exists a (static) PLS (call it PLS\_t) for $f$ on anonymous rings. For PLS\_t, denote the marker by $M_t$, the verifier at node $v$ by $V_t(v)$, and the label of node $v$ by $L_t(v)$. Let us construct a (static) PLS (called PLS\_s) for the predicate (the size of the ring is $n$) using PLS\_t as a procedure. For PLS\_s, denote the marker by $M_s$, the verifier at node $v$ by $V_s(v)$, and the label of node $v$ by $L_s(v)$. Each label $L_s(v)$ contains three fields. $L_s(v) = (L_{s_1}(v), L_{s_2}(v), L_{s_3}(v))$. Field $L_{s_1}(v)$ is Boolean: $1$ indicates that node $v$ has a virtual token and $0$ otherwise. Field $L_{s_2}(v)$ is an integer which indicates the clockwise distance of node $v$ from the token holder. Field $L_{s_3}(v) = L_t(v)$ is the label generated by $M_t$ for the virtual token in the $L_{s_1}(v)$ label. Recall that for a (static) PLS, the marker is not necessarily distributed.

\begin{algorithm}
\caption{Code of $M_s$}
\label{alg:ms}
\begin{algorithmic}[1]
\STATE Choose a node $v$ and set $L_{s_1}(v) \Leftarrow 1$, $L_{s_2}(v) \Leftarrow 0$
\STATE For any node $u \neq v$ which is $k$ hops away clockwise from $v$, set $L_{s_1}(u) \Leftarrow 0$, $L_{s_2}(u) \Leftarrow k$
\STATE Call $M_t$ with $Ls_1$ as input, for any node $w$, set $L_{s_3}(w) \Leftarrow L_{t}(w)$
\end{algorithmic}
\end{algorithm}

\begin{algorithm}
\caption{Code of $V_s$ for Node $v$}
\label{alg:vs}
\begin{algorithmic}[1]
\STATE /*Denote $v$'s predecessor by $u$, and $v$'s successor by $w$.*/
\STATE Send $L_{s_1}(v)$, $L_t(v)$, $L_t(u)$ and $L_t(w)$ as input to $V_t(v)$
\STATE \textbf{IF} any of the following constraints is violated, output 0
\begin{ALC@g}
\STATE C1: $V_t(v) = 1$
\STATE C2: $(L_{s_1}(v) = 1 \land L_{s_2}(v) = 0 \land L_{s_2}(u) = n - 1) \lor (L_{s_1}(v) = 0 \land L_{s_2}(v) = L_{s_2}(u) + 1)$
\end{ALC@g}
\STATE \textbf{ELSE} output 1
\end{algorithmic}
\end{algorithm}

Let us prove that PLS\_s (see \autoref{alg:ms} and \autoref{alg:vs}) is correct.\\ \textbf{Case 1:} The size of the ring is indeed $n$. For $k=0$,$1$,..., $n-1$, let node $v_{k+1 \bmod n}$ be the successor of $v_k$. First, the marker $M_s$ labels each node such that there is only one virtual token. For every node $v_k$, verifier $V_t(v_k)$ outputs $1$ due to the correctness of PLS\_t, which means C1 is satisfied (\autoref{alg:vs}). WLOG, let $v_0$ be the token holder. Line 1-2 of \autoref{alg:ms} ensures $(L_{s_1}(v_0) = 1 \land L_{s_2}(v_0) = 0 \land L_{s_2}(v_{n-1}) = n - 1)$ and $(L_{s_1}(v_k) = 0 \land L_{s_2}(v_k) = L_{s_2}(v_{k-1}) + 1)$ for every node $v_k \neq v_0$. Therefore, C2 (\autoref{alg:vs}) is satisfied for every node.\\
\textbf{Case 2:} The size of the ring is $X \neq n$. For $k=0$,$1$,..., $X-1$, let node $v_{k+1 \bmod X}$ be the successor of $v_k$. Assume for contradiction that $V_s(v_k)$ outputs 1 for every node $v_k$. No $V_t(v_k)$ outputs 0 (otherwise, $V_s(v_k)$ would have output 0 due to violation of C1 of \autoref{alg:vs}). By the correctness of the PLS\_t, there is only one virtual token holder (WLOG, let it be $v_0$). Since C2 is satisfied by our assumption, every node $v_k \neq v_0$ has $L_{s_2}(v_k) = L_{s_2}(v_{k-1}) + 1$. In addition, we have $L_{s_2}(v_0) = 0$. Therefore, we have $L_{s_2}(v_{X-1}) = L_{s_2}(v_{X-2}) + 1 = L_{s_2}(v_{X-3}) + 2 =...=L_{s_2}(v_{0}) + X - 1 = X - 1 \neq n - 1$. It violates C2 of $V_s(v_0)$. It is a contradiction.
\end{sketchproof}
\end{hide}

\begin{lemma}[\protect{Lemma 3.1 in \cite{korman2010proof}}]
\label{lem:nopls}
There is no (static) PLS that verifies whether the size of an anonymous ring is $n$.
\end{lemma}



\begin{extra}
The proof of these lemmas, as well as the other omitted proofs, can be found in the longer version of this extended abstract.  
\end{extra}
Combining \autoref{lem:reduction1}, \autoref{lem:reduction2}, and \autoref{lem:nopls} completes the proof that there does not exist an RPLS for token passing on anonymous rings without some additional assumption such as the one we made that the size of the anonymous ring is known to every node.
\section{RPLS for Token Passing on General Graphs $G$}
\label{sec:general}
In this section, we assume that each node has a unique identifier. We also assume that each node knows the number $n$ of nodes in the graph.
\begin{definition}
\label{def:checkpoint}
A round $r$ is a \textbf{checkpoint} if $r \bmod n = 0$.
\end{definition}
Informally, the first idea behind the RPLS for general graphs was to try to use the PLS of \cite{korman2010proof} for a {\em static} rooted spanning tree. That is, as long as the token resides in some single node $v$ and does not move out of $v$, let $v$ be the root and verify that $v$ holds a token. Every other node $u\neq v$ then needs to verify it does not hold a token, and (using the PLS) that a root does exist.
Unfortunately, consider the case that the token is passed from $v$ to some other node $w$ over an edge that does {\em not} belong to the tree. While it may be easy to update the tree so that $w$ is the new root, it may take diameter (of the tree) time for a distributed marker to update the PLS of \cite{korman2010proof}.

Hence, the next idea is to have a second part of the RPLS to verify the movements of the token at any round in the execution after $r_0$. Whenever a node $u$ passes the token to some other node $w$, both $u$ and $w$ record this move, as well as the round number when this happened. Thus, had we allowed them to remember all the history, they could have simulated their actions in the execution. As is shown later by induction, if the records of all the nodes match those of their neighbors, and match the assumption that at time $r_0$, the token holder was $v$, then there exists indeed a single token. 

Unfortunately, the above method would have used unbounded history, as well as an unbounded round number. The main new idea is how to truncate the history from time to time. Specifically, at each round $r$ that is {\em not} a checkpoint, let the last checkpoint be called $r_2$.  The static tree is one that verifies the place of the token at the {\em one before last} checkpoint $r_1$. Any part of the history before $r_1$ is forgotten and the count of the time starts by setting $r_1=0$. The static tree is replaced every checkpoint, so in the example of $r$ above, it will be replaced at the next checkpoint $r_3$. Informally, that future static tree will
prove the location of the token in the {\em previous} checkpoint $r_2$ ($n$ rounds earlier) since it takes a diameter time to construct such a PLS.  That future static tree is under construction at $r_2 >r> r_3$ and is called then the candidate tree. At time $r_3$, all the history information regarding rounds earlier than $r_2$ is forgotten. 

The RPLS maintains the following variables for each node $v$:

\textbullet $static\_root(v)$ --- id of the root node of the static tree.

\textbullet $static\_parent(v)$ --- $NULL$ or the id of $v$'s parent on the static tree.

\textbullet $static\_dist(v)$ --- distance from the root of the static tree.

\textbullet $cand\_root(v)$ --- id of the root of the candidate tree.

\textbullet $cand\_dist(v)$ --- distance from the root of the candidate tree.

\textbullet $cand\_parent(v)$ --- $NULL$ or the id of $v$'s parent on the candidate tree.

\textbullet $dynamic\_parent(v)$ --- $NULL$ or the id of $v$'s parent on the dynamic tree.

\textbullet $token\_in(v)$ --- set of pairs of $(r, id(u))$ which logs the round $r$ at which the token was sent from $v$'s neighbor $u$ to $v$.

\textbullet $token\_out(v)$ --- set of pairs of $(r, id(u))$ which logs the round $r$ at which the token was sent from $v$ to $v$'s neighbor $u$. 

\textbullet $L(v)$ is a collection of the local variables of $v$ that are listed above. 

\begin{definition}
The \textbf{static tree} of round $r$ is the collection of edges pointed by the $static\_parent_r(v)$ pointers at each node $v$. The \textbf{dynamic tree} of round $r$ is the collection of edges pointed by the $dynamic\_parent_r(v)$ pointers at each node $v$. The \textbf{candidate tree} between checkpoint $r$ (including) and checkpoint $r+n$ (excluding) is the dynamic tree of checkpoint $r$.
\end{definition}

\noindent \textbf{Specification of the legal initial labeling: }
\begin{hide}
Assume that at the end of some checkpoint $r_0$, there is exactly one token holder $v$ with $s_{r_0}(v) = 1$ while any node $w \neq v$ has $s_{r_0}(w) = 0$.
\end{hide}
Select any spanning tree rooted at the token holder $v$ and set $static\_dist_{r_0}(v) \Leftarrow 0$, $static\_parent_{r_0}(v) \Leftarrow NULL$. Then $\forall w \neq v$ that is $k$ hops away from $v$, denote $w$'s parent by some neighbor $p$ that is $k-1$ hops away from $v$. Also, set $static\_dist_{r_0}(w) \Leftarrow k$, $static\_parent_{r_0}(w) \Leftarrow id(p)$. In addition, for every $u \in G$, $static\_root_{r_0}(u) \Leftarrow id(v)$, $cand\_root_{r_0}(u) \Leftarrow id(v)$, $cand\_parent_{r_0}(u) \Leftarrow static\_parent_{r_0}(u)$, $cand\_dist_{r_0}(u) \Leftarrow static\_dist_{r_0}(u)$, and $dynamic\_parent_{r_0}(u)$ $\Leftarrow static\_parent_{r_0}(u)$. At last, set $token\_in_{r_0}(u) \Leftarrow \emptyset$, $token\_out_{r_0}(u) \Leftarrow \emptyset$.

See the RPLS in \autoref{alg:generalverifier} and \autoref{alg:generalmarker}. The entries in $token\_in(v)$ and $token\_out(v)$ are sorted in a linearly increasing order over the rounds of the entries. Denote the $k^{th}$ entries of $token\_in(v)$ and $token\_out(v)$ by $en_k^{in}$ and $en_k^{out}$ respectively. Let $\lvert token\_in(v)\rvert = a$ and $\lvert token\_out(v)\rvert = b$. Given any $(r_1, id(u))$ and $(r_2, id(w))$, define $(r_1, id(u)) \prec (r_2, id(w))$ if $r_1 < r_2$. 

Note that for simplicity of exposition, the round $r$ looks unbounded in \autoref{alg:generalverifier} and \autoref{alg:generalmarker}. However, a ``bounded timestamp" can be easily implemented by encoding the rounds modulo $2n$, since the history before the previous checkpoint is forgotten.
\begin{algorithm}
\caption{$\mathcal{V}(v)$ in round $r$}
\label{alg:generalverifier}
\begin{algorithmic}[1]
\STATE \textbf{IF} $checkST(v) = 0$, output 0   /* see \autoref{alg:checkst}*/
\STATE \textbf{ELSE IF} any of the following conditions is violated, output 0:
    \begin{ALC@g}
    \STATE /*H0-H3 check consistency of the history logs*/
    \STATE /*Denote checkpoint $r - 1 - ((r-1) \bmod n) - n$ by $r_{c_1}$*/
    \STATE H0: $\forall (r_k, id(u)) \in token\_out_{r-1}(v)$ or $token\_in_{r-1}(v)$, $r_{c_1} < r_k < r \land u \in N(v)$
    \STATE H1: $\forall (r_1, id(u)), (r_2, id(w)) \in token\_in_{r-1}(v) \lor token\_out_{r-1}(v), r_1 \neq r_2$
    \STATE H2.1: if $static\_root_{r-1}(v) = id(v) \land s_{r-1}(v)=0$, then $b = a + 1>0$ and $en_1^{out} \prec en_1^{in} \prec en_2^{out} \prec en_2^{in} \prec ...\prec en_a^{out} \prec en_a^{in} \prec en_{a+1}^{out}$
    \STATE H2.2: if $static\_root_{r-1}(v) \neq id(v) \land s_{r-1}(v)=1$, then $a = b + 1>0$ and $en_1^{in} \prec en_1^{out} \prec en_2^{in} \prec en_2^{out} \prec ...\prec en_b^{in} \prec en_b^{out} \prec en_{b+1}^{in}$;
    \STATE H2.3: if $static\_root_{r-1}(v) = id(v) \land s_{r-1}(v)=1$, then $a = b$ and $en_1^{out} \prec en_1^{in} \prec en_2^{out} \prec en_2^{in} \prec ...\prec en_b^{out} \prec en_b^{in}$
    \STATE H2.4: if $static\_root_{r-1}(v) \neq id(v) \land s_{r-1}(v)=0$, then $a = b$ and $en_1^{in} \prec en_1^{out} \prec en_2^{in} \prec en_2^{out} \prec ...\prec en_a^{in} \prec en_a^{out}$
    \STATE H3.1: $\forall u \in N(v)$, if $\exists (r, id(u)) \in token\_out_{r-1}(v)$, then\\ $\exists (r, id(v)) \in token\_in_{r-1}(u)$
    \STATE H3.2: $\forall u \in N(v)$, if $\exists (r, id(u)) \in token\_in_{r-1}(v)$, then\\ $\exists (r, id(v)) \in token\_out_{r-1}(u)$
    \end{ALC@g}
\STATE \textbf{ELSE} output 1
\end{algorithmic}
\end{algorithm}

\begin{algorithm}[H]
\caption{$\mathcal{M}(v)$ in round $r$}
\label{alg:generalmarker}
\begin{algorithmic}[1]
\STATE \textbf{IF} $cand\_parent(v) \neq NULL$ $\land$ $cand\_dist_{r-1}(cand\_parent(v)) \neq NULL$:
    \begin{ALC@g}
    \STATE $cand\_root(v) \Leftarrow cand\_root_{r-1}(cand\_parent(v))$
    \STATE $cand\_dist(v) \Leftarrow cand\_dist_{r-1}(cand\_parent(v)) + 1$
    \end{ALC@g}
\STATE \textbf{IF} token is passed from $u \in N(v)$ to $v$ in $r$:
    \begin{ALC@g}
    \STATE $dynamic\_parent(v) \Leftarrow NULL$, add $(r, id(u))$ in $token\_in(v)$
    \end{ALC@g}
\STATE \textbf{ELSE IF} token is passed from $v$ to $u \in N(v)$ in $r$:
    \begin{ALC@g}
    \STATE $dynamic\_parent(v) \Leftarrow id(u)$, add $(r, id(u))$ in $token\_out(v)$
    \end{ALC@g}
\STATE \textbf{IF} $r\bmod{n} = 0$:
    \begin{ALC@g}
    \STATE remove all the entries with $r_k \leq r-n$ in $token\_out(v)$ and $token\_in(v)$.
    \STATE $static\_root(v) \Leftarrow cand\_root(v)$
    \STATE $static\_parent(v) \Leftarrow cand\_parent(v)$
    \STATE $static\_dist(v) \Leftarrow cand\_dist(v)$
    \STATE $cand\_parent(v) \Leftarrow dynamic\_parent(v)$
    \STATE \textbf{IF} $s(v) = 1$:
        \begin{ALC@g}
        \STATE $cand\_root(v) \Leftarrow id(v)$, $cand\_dist(v) \Leftarrow 0 $
        \end{ALC@g}
    \STATE \textbf{ELSE} 
        $cand\_root(v) \Leftarrow NULL$, $cand\_dist(v) \Leftarrow NULL$
    \end{ALC@g}
\end{algorithmic}
\end{algorithm}

\begin{algorithm} [H]
\caption{Procedure $checkST(v)$ in round $r$}
\label{alg:checkst}
\begin{algorithmic}[1]
\STATE \textbf{IF} any of the following conditions is violated, return 0:
    \begin{ALC@g}
    \STATE /*S1-S3 check the correctness of the static spanning tree*/
    \STATE S1: $\forall u \in N(v)$: $static\_root_{r-1}(v) = static\_root_{r-1}(u)$
    \STATE S2: if $static\_root_{r-1}(v) =id(v)$, then\\ $static\_parent_{r-1}(v) = NULL \land static\_dist_{r-1}(v) = 0$
    \STATE S3: if $static\_root_{r-1}(v) \neq id(v)$, then $\exists u \in N(v): static\_parent_{r-1}(v) = id(u) \land static\_dist_{r-1}(v) = static\_dist_{r-1}(u) + 1$
    \end{ALC@g}
\STATE \textbf{ELSE} return 1
\end{algorithmic}
\end{algorithm}
\begin{hide}
\subsection{Proof of Correctness}
\begin{lemma}
\label{lem:initial}
The specification of the legal initial labeling indeed gives a legal initial labeling.
\end{lemma}

\begin{sketchproof}
By line 1 of \autoref{alg:generalverifier}, $\mathcal{V}(v)$ calls procedure $checkST(v)$ (\autoref{alg:checkst}) which is a (static) PLS given in \cite{korman2010proof} that verifies whether a spanning tree is rooted at some node.  The specified labeling ensures that a static (spanning) tree is rooted at $v$ and $checkST(v)$ returns 1. Since $token\_in_{r_0}(v) = \emptyset$ and $token\_out_{r_0}(v) = \emptyset$, H0-H3 of \autoref{alg:generalverifier} are also satisfied. All the verifiers output 1 in $r_0+1$.
\end{sketchproof}

Under the assumption that the RPLS is executed with $Alg$ and with a legal initial labeling at time $r_0$, and the execution is correct until $r+1$ (excluding $r+1$), we prove the following lemmas (\autoref{lem:dynamictree}, \autoref{lem:treeupdate} and \autoref{lem:historylog}).

\begin{lemma}
\label{lem:dynamictree}
The collection of variables $dynamic\_parent_{r_k}(u)$ for every node $u$ is a spanning tree rooted at the token holder at round $r_k$ ($r_0 \leq r_k \leq r$).
\end{lemma}

\begin{sketchproof}
By \autoref{lem:initial}, the static tree for $r_0$ is rooted at the token holder of $r_0$. Since we have $dynamic\_parent_{r_0}(u) \Leftarrow static\_parent_{r_0}(u)$ for every node $u$ by the specification of the legal initial labeling, the dynamic tree is the same as the static tree in $r_0$. Assume by induction, that the dynamic tree is a spanning tree rooted at the token holder (denoted by $v$) of $r_k-1$ ($r_k-1 \geq r_0$). If the token is not passed in $r_k$, then clearly no label changes, so the correctness for $r_k$ follows from the correctness for $r_k-1$. If the token is passed in $r_k$ from $v$ to a neighbor $u$, $\mathcal{M}(v)$ sets $dynamic\_parent_{r_k}(v) \Leftarrow id(u)$ by line 7 of \autoref{alg:generalmarker} and $\mathcal{M}(u)$ sets $dynamic\_parent_{r_k}(u) \Leftarrow NULL$ by line 5 of \autoref{alg:generalmarker}. For any node $w \neq v$ or $u$, $dynamic\_parent(w)$ remains unchanged. Clearly, the collection of $dynamic\_parent_{r_k}(p)$ forms a spanning tree which is rooted at the token holder $u$ of $r_k$.
\end{sketchproof}

\begin{lemma}
\label{lem:treeupdate}
Constraints S1-S3 of \autoref{alg:checkst} are satisfied for every $u$ when $\mathcal{V}(u)$ checks them in $r+1$.
\end{lemma}

\begin{sketchproof}
By \autoref{lem:initial}, constraints S1-S3 of \autoref{alg:checkst} are satisfied for every $u$ when $\mathcal{V}(u)$ checks them in $r_0+1$. Induction hypothesis: constraints S1-S3 of \autoref{alg:checkst} are satisfied for every $u$ when $\mathcal{V}(u)$ checks them in $r$. \\
\textbf{Case 1:} round $r$ is not a checkpoint. No changes are made to the static tree in $r$, so the correctness for $r+1$ follows from the correctness for $r$.\\
\textbf{Case 2:} round $r$ is a checkpoint. By \autoref{lem:dynamictree}, we know that the dynamic tree of checkpoint $r-n$ is rooted at the token holder (denote it by $v$) of $r-n$. By line 13 of \autoref{alg:generalmarker}, for any node $u$, $cand\_parent_{r-n}(u)=dynamic\_parent_{r-n}(u)$ which makes the candidate tree (call it $T_{cand}$) between $r-n$ and $r$ the same as the dynamic tree of $r-n$. In addition, by line 14-16 of \autoref{alg:generalmarker}, $v$ sets itself as the root of $T_{cand}$ ($cand\_root_{r-n}(v) = id(v)$) and $cand\_dist_{r-n}(v) = 0$ while any other node $w \neq v$ sets $cand\_root_{r-n}(w)=NULL$ and $cand\_dist_{r-n}(w)=NULL$. Then in any round $r-n+k$ ($k<n$) that follows checkpoint $r-n$, by line 1-3 of \autoref{alg:generalmarker}, every node $u \neq v$ that is $k$ hops away from $v$ on $T_{cand}$ sets $cand\_root(u)=id(v)$ and $cand\_dist(u)=k$. By the end of round $r-1$, every node $u \neq v$ has $cand\_root(u)=id(v)$ and its distance to $v$ on $T_{cand}$ encoded in $cand\_dist(u)$ since the diameter of the system is bounded by $n-1$. By line 10-12 of \autoref{alg:generalmarker}, a new static tree is created by copying $T_{cand}$. Clearly, this new static tree is rooted at $v$ with all the labels satisfying S1-S3 in $r+1$.
\end{sketchproof}

\begin{lemma}
\label{lem:historylog}
Constraints H0-H3 of \autoref{alg:generalverifier} are satisfied for every $u$ when $\mathcal{V}(u)$ checks them in $r+1$.
\end{lemma}

\begin{sketchproof}
By \autoref{lem:initial}, constraints H0-H3 of \autoref{alg:generalverifier} are satisfied for every $u$ when $\mathcal{V}(u)$ checks them in $r_0+1$. Induction hypothesis: H0-H3 are satisfied for every $u$ when $\mathcal{V}(u)$ checks them in $r$ ($r \geq r_0+1$). \\
\textbf{Case 1:} round $r$ is not a checkpoint. If the token is not passed in $r$, then no label changes and the correctness for $r+1$ follows from the correctness for $r$. If the token is passed in $r$ from $v$ to a neighbor $u$, TPA sets $s_r(v)=0$ and $s_r(u)=1$. By line 4-7 of \autoref{alg:generalmarker}, an entry $(r, id(u))$ is added to $token\_out_r(v)$ and an entry $(r, id(v))$ is added to $token\_in_r(u)$. For every node $w \neq v$ and $w \neq u$, no new entries are added to $L_r(w)$ and the correctness for $r+1$ follows from the correctness for $r$. For $u$ and $v$, clearly, H0, H1 and H3 are still satisfied in $r+1$. Now let us examine constraint H2 for $v$ and $u$. \\
For $v$, there are two cases: \\
(1) $static\_root_{r-1}(v) = id(v) \land s_{r-1}(v)=1$ (node $v$ was the static root and $v$ was the token holder in $r-1$). By the induction hypothesis, the condition of H2.3 is satisfied in $r$. The size of $token\_in_{r-1}(v)$ and $token\_out_{r-1}(v)$ is equal. The static root remains the same in $r$ compared to $r-1$ (since $r$ is not a checkpoint). Since the token was passed from $v$ to $u$ in $r$, the interface variable $s_r(v) = 0$. With the new entry $(r, id(u))$ added to $token\_out_r(v)$, there is one more entry in $token\_out_r(v)$ than $token\_in_r(v)$ and the order of the entries in them clearly satisfies H2.1 in $r+1$.\\
(2) $static\_root_{r-1}(v) \neq id(v) \land s_{r-1}(v)=1$ (node $v$ was not the static root and $v$ was the token holder in $r-1$). By the induction hypothesis, the condition of H2.2 is satisfied in $r$. There is one more entry in $token\_in_{r-1}(v)$ than $token\_out_{r-1}(v)$. The static root remains the same in $r$ compared to $r-1$. Since the token was passed from $v$ to $u$ in $r$, the interface variable $s_r(v) = 0$. With the new entry $(r, id(u))$ added to $token\_out_r(v)$, the size of $token\_in_r(v)$ and $token\_out_r(v)$ is equal and the order of the entries in them clearly satisfies H2.4 in $r+1$.\\
For $u$, there are two cases: \\
(1) $static\_root_{r-1}(u) = id(u) \land s_{r-1}(u)=0$ (node $u$ was the static root and $u$ was not the token holder in $r-1$). By the induction hypothesis, condition H2.1 is satisfied in $r$. There is one more entry in $token\_out_{r-1}(u)$ than in $token\_in_{r-1}(u)$. The static root remains the same in $r$ compared to $r-1$. Since the token was passed from $v$ to $u$ in $r$, the interface variable $s_r(u) = 1$. With the new entry $(r, id(v))$ added to $token\_in_r(u)$, the size of $token\_in_r(u)$ and $token\_out_r(u)$ is equal and the order of the entries in them clearly satisfies H2.3 in $r+1$.\\
(2) $static\_root_{r-1}(u) \neq id(u) \land s_{r-1}(u)=0$ (node $u$ was not the static root and $u$ was not the token holder in $r-1$). By the induction hypothesis, the condition of H2.4 is satisfied in $r$. The size of $token\_in_{r-1}(u)$ and $token\_out_{r-1}(u)$ is equal. The static root remains the same in $r$ compared to $r-1$. Since the token was passed from $v$ to $u$ in $r$, the interface variable $s_r(u) = 1$. With the new entry $(r, id(v))$ added to $token\_in_r(u)$, there is one more entry in $token\_in_r(u)$ than $token\_out_r(u)$ and the order of the entries in them clearly satisfies H2.2 in $r+1$.\\
\textbf{Case 2:} round $r$ is a checkpoint. By induction over the rounds in the case 1 above, we know that $token\_in_{r-1}(u)$ and $token\_out_{r-1}(u)$ for every node $u$ contain all the entries for the token movements from the previous checkpoint $r-n$ (including) until $r$ (excluding). By line 4-7 of \autoref{alg:generalmarker}, an additional entry is added to $token\_in_r(u)$ or $token\_out_r(u)$ in $r$ if a token is passed to $u$ or from $u$ in $r$. By line 9 of \autoref{alg:generalmarker}, the entries of any history log before $r - n$ (including) are removed. Note that by line 10-12 of \autoref{alg:generalmarker}, the static tree at the end of $r$ is replaced by the tree (denote it by $T_{cand}$ and its root by $v$) that was the candidate tree between checkpoints $r-n$ and $r$. By line 13 of \autoref{alg:generalmarker}, this candidate tree $T_{cand}$ was formed by copying the dynamic tree of checkpoint $r-n$ which was rooted at the token holder at the end of $r-n$. Therefore, node $v$ was the token holder at $r-n$.
\begin{observation}
for every node $u$, $token\_in_r(u)$ and $token\_out_r(u)$ only contain entries $r-n$ (excluding) and $r$ (including $r$). Every movement of the token during this period where it is passed from $u$ to a neighbor $w \in N(u)$ is saved as an entry in $token\_out_r(u)$ and an entry in $token\_in_r(w)$.
\end{observation}
Since the execution is correct until $r+1$, clearly, H0, H1 and H3 are satisfied in $r+1$. Now let us examine H2 of any node $u$ at the end of round $r$. There are 4 cases for $u$: \\
(1) $u = v$ ($u$ is the root of $T_{cand}$) and $s_r(u) = 0$. Since the execution is correct until $r+1$, node $u$ was the only token holder at round $r-n$ but it does not hold the token at the end of $r$ ($s_r(u)=0$). The token must have been passed from $u$ for the first time in some round $r-n < r_1 \leq r$ to a neighbor $w \in N(u)$, entry $en_1^{out} = (r_1, id(w))$ is saved in $token\_out_r(u)$ and $en_1^{out}$ precedes any other entry (if such exists) in $token\_in_r(u)$ and $token\_out_r(u)$. After $r_1$, each time the token was passed back to $u$, it must have been passed from $u$ to a neighbor afterwards (otherwise $u$ would still be the token holder in $r$). These movements are saved as entries in $token\_in_r(u)$ and $token\_out_r(u)$ chronologically and it is clear that there is one more entry in $token\_out_r(u)$ than $token\_in_r(u)$, and the order of the entries in them follow constraint H2.1.\\
(2) $u = v$ ($u$ is the root of $T_{cand}$) and $s_r(u) = 1$. Since the execution is correct until $r+1$, node $u$ was the only token holder at round $r-n$ but it holds the token at the end of $r$ ($s_r(u)=0$). If the token was never passed from $u$ staring from $r-n$, then both $token\_out_r(u)$ and $token\_in_r(u)$ are empty sets. H2.3 follows. If the token was ever passed from $u$ to a neighbor, then each time it happens, it must have been passed back to $u$ afterwards (otherwise $u$ would not be the token holder in $r$). These movements are saved as entries in $token\_in_r(u)$ and $token\_out_r(u)$ chronologically and it is clear that there are the same amount of entries in $token\_out_r(u)$ and $token\_in_r(u)$, and the order of the entries in them follow constraint H2.3.\\
(3) $u \neq v$ ($u$ is not the root of $T_{cand}$) and $s_r(u) = 0$. Since the execution is correct until $r+1$, node $u$ was not the token holder at round $r-n$ and it does not hold the token at the end of $r$ ($s_r(u)=0$). If the token was never passed to $u$ staring from $r-n$, then both $token\_out_r(u)$ and $token\_in_r(u)$ are empty sets. H2.4 follows. If the token was passed to $u$, then each time it happens, the token must have been passed from $u$ to a neighbor afterwards (otherwise $u$ would still be the token holder in $r$). These movements are saved as entries in $token\_in_r(u)$ and $token\_out_r(u)$ chronologically and it is clear that there are the same amount of entries in $token\_out_r(u)$ and $token\_in_r(u)$, and the order of the entries in them follow constraint H2.4.\\
(4) $u \neq v$ ($u$ is not the root of $T_{cand}$) and $s_r(u) = 1$. Since the execution is correct until $r+1$, node $u$ was not the token holder at round $r-n$ and it holds the token at the end of $r$ ($s_r(u)=0$). The token must have been passed to $u$ for the first time in some round $r-n < r_1 \leq r$ to a neighbor $w \in N(u)$. After $r_1$, each time the token is passed from $u$ to a neighbor, it must have been passed back to $u$ afterwards (otherwise $u$ would not be the token holder in $r$). These movements are saved as entries in $token\_in_r(u)$ and $token\_out_r(u)$ chronologically and it is clear that there is one more entry in $token\_in_r(u)$ than $token\_out_r(u)$, and the order of the entries in them follow constraint H2.2.
\end{sketchproof}

\begin{sketchproof}[thm:correct]
By \autoref{lem:initial}, \autoref{lem:treeupdate} and \autoref{lem:historylog}, the theorem follows.
\end{sketchproof}

\begin{sketchproof}[thm:incorrect]
In this proof, S1-S3 refer to S1-S3 in \autoref{alg:checkst} and H0-H3 refer to H0-H3 in \autoref{alg:generalverifier}. An entry $en$ \emph{precedes} an entry $en'$ if $en \prec en'$. For any node $u$, an entry $en^{in} \in token\_in(u)$ \emph{directly precedes} an entry $en^{out} \in token\_out(u)$ if there does not exist $en' \in token\_in(u)$ such that $en^{in} \prec en' \prec en^{out}$. For the identifier of any node $u$, we might just write $u$ instead of $id(u)$ for ease of reading. Assume for contradiction, that all the verifiers output 1 in round $r+1$ which implies that S1-S3 and H0-H3 are satisfied in $r+1$. \\
\textbf{Case 1:} No token exists in $r$. By S1-S3, there is a unique node $p_0$ such that for every node $u$, $static\_root_r(u) = id(p_0)$. Since no token exists in $r$, node $u$ has $s_r(u) = 0$. \\
Induction base: By $b=a+1>0$ of H2.1, $p_0$ must have at least one entry $(r_1, p_1)$ in $token\_out_r(p_0)$ such that $r_1 > r_{c_1}$ and $p_1 \in N(p_0)$. \\
Induction step: for any node $p_k$, if there exists entry $(r_{k+1}, p_{k+1}) \in token\_out_r(p_k)$ such that $r_{k+1} > r_{c_1}$ and $p_{k+1} \in N(p_k)$, then there exists a round $r_{k+2} > r_{k+1}$, a node $p_{k+2} \in N(p_{k+1})$ and an entry $(r_{k+2}, p_{k+2}) \in token\_out_r(p_{k+1})$.

This induction directly follows from the assumption that H2-H3 are satisfied in $r+1$: H3.1 ensures that there exists an entry $(r_{k+1}, p_k) \in token\_in_r(p_{k+1})$. H2.1 and H2.4 (note that $s_r(u) = 0$ for every node $u$) ensure that $(r_{k+1}, p_k) \in token\_in_r(p_{k+1})$ must directly precede an entry $(r_{k+2}, p_{k+2}) \in token\_out_r(p_{k+1})$ such that $r_{k+2} > r_{k+1}$ and $p_{k+2} \in N(p_{k+1})$.

By induction, there exists a node $p_i$ such that there exists an entry $(r_{i+1}, p_{i+1}) \in token\_out_r(p_i)$ ($p_{i+1} \in N(p_i)$) such that $r_{i+1} > r$. It is a contradiction to H0 since the round number in any entry of $token\_out_r(p)$ must be smaller than $r+1$.\\
\textbf{Case 2:} Multiple (more than 1) tokens in round $r$. By S1-S3, there is a unique node $v$ such that for every node $p$, $static\_root_r(p) = id(v)$. Since there are multiple token holders in $r$, at most one of them is node $v$. First, let us look at any token holder $u$ such that $u \neq v$. We have $s_r(u) = 1$.\\
Induction base: By $a=b+1>0$ of H2.2, $u$ must have at least one entry in $token\_in_r(u)$. Let the last entry (no other entry is preceded by it) of $token\_in_r(u)$ be $(r_i, x_{i-1})$ such that $r_{c_1} < r_i \leq r$ and $x_{i-1} \in N(u)$. \\
Induction step: for any node $x_i$, if there exists entry $(r_i, x_{i-1}) \in token\_in_r(x_i)$ such that $r_{c_1} < r_i \leq r$ and $x_{i-1} \in N(x_i)$, then there exists an entry $(r_i, x_i) \in token\_out_r(x_{i-1})$ (by H3) and one of the following holds (by H2): \\(1)there exists a round $r_{i-1}$ ($r_{c_1} < r_{i-1} < r_i$), a node $x_{i-2} \in N(x_{i-1})$, and entry $(r_{i-1}, x_{i-2}) \in token\_in_r(x_{i-1})$ which directly precedes $(r_i, x_i) \in token\_out_r(x_{i-1})$. \\(2) $x_{i-1} = v$ and $(r_i, x_i) \in token\_out_r(v)$ precedes all the other entries (if such exist) of $token\_out_r(v)$ and $token\_in_r(v)$.

The induction directly follows from the assumption that H2-H3 are satisfied in $r+1$. By this induction, we find a chain of entries. This chain indicates how the current token in $u$ was passed from $v$ through $k$ nodes to $u$: $(r_0,p_0) \in token\_out_r(v)$, $(r_0,v) \in token\_in_r(p_0)$, $(r_1, p_1) \in token\_out_r(p_0)$, $(r_1, p_0) \in token\_in_r(p_1)$,..., $(r_{k}, u) \in token\_out_r(p_{k-1})$, $(r_{k}, p_{k-1}) \in token\_in_r(u)$. In this chain, $(r_0,p_0) \in token\_out_r(v)$ is not preceded by any other entry of $token\_out_r(v)$ and $token\_in_r(v)$, which indicates that the token remained in $v$ from $r_{c_1}$ and it was passed from $v$ to $p_0 \in N(v)$ in round $r_0$. Consider $v$ as $p_{-1}$ and $u$ as $p_k$. For any node $p_i$ ($0 \leq i < k$), entry $(r_i, p_{i-1}) \in token\_in_r(p_i)$ directly precedes the next entry $(r_{i+1}, p_{i+1}) \in token\_out_r(p_i)$ on this chain. The two consecutive entries indicate that the token was passed from $p_{i-1}$ to $p_i$ at round $r_i$ and the token stayed at $p_i$ until it was passed to $p_{i+1}$ at round $r_{i+1}$. Entry $(r_k, p_{k-1})$ is the last entry in $token\_in_r(u)$ and it indicates that the current token at $u$ was passed to $u$ at round $r_k$ and stayed at $u$ until $r$ (including).

Now let us examine any other token holder $w \neq u$. If $w \neq v$, then clearly the induction also applies for $w$. If $w = v$, from the above chain of entries, we know that there exists at least one entry in $token\_out_r(v)$. By $a=b$ in H2.3, we know that there exists at least one entry in $token\_in_r(v)$. Therefore, the above induction for $u \neq v$ also applies for $v$($w$).

As a result, we find a chain of entries for $w \neq u$ using the same induction which indicates how the current token in $w$ was passed from $v$ through $j$ nodes to $w$: $(r'_0,q_0) \in token\_out_r(v)$, $(r'_0,v) \in token\_in_r(q_0)$, $(r'_1, q_1) \in token\_out_r(q_0)$, ..., $(r'_{j}, w) \in token\_out_r(q_{j-1})$, $(r'_{j}, q_{j-1}) \in token\_in(w)$. In this chain, $(r'_0,q_0) \in token\_out_r(v)$ is not preceded by any entry of $token\_out_r(v)$ and $token\_in_r(v)$, which indicates that the token remained in $v$ from $r_{c_1}$ and it was passed from $v$ to $q_0 \in N(v)$ in round $r'_0$. Consider $v$ as $q_{-1}$ and $w$ as $q_j$. For any node $q_i$ ($0 \leq i < j$), entry $(r'_i, q_{i-1}) \in token\_in_r(q_i)$ directly precedes the next entry $(r'_{i+1}, q_{i+1}) \in token\_out_r(q_i)$ on this chain. The two consecutive entries indicate that the token was passed from $q_{i-1}$ to $q_i$ at round $r'_i$ and the token stayed at $q_i$ until it was passed to $q_{i+1}$ at round $r'_{i+1}$. Entry $(r'_j, q_{j-1})$ is the last entry in $token\_in_r(w)$ and it indicates that the current token at $w$ was passed to $w$ at round $r'_j$ and stayed at $w$ until $r$ (including). 

Now let us compare the two chains.
\begin{observation}
\label{obs:first}
Neither $(r_0,p_0) \in token\_out_r(v)$ nor $(r'_0,q_0) \in token\_out_r(v)$ is preceded by any entry in $token\_in_r(v)$.\\
\end{observation}

\begin{observation}
\label{obs:second}
For $0 \leq i < k$, entry $(r_{i+1},p_{i+1}) \in token\_out_r(p_i)$ is \emph{directly preceded} by entry $(r_i,p_{i-1}) \in token\_in_r(p_i)$.\\
For $0 \leq i < j$, entry $(r'_{i+1},q_{i+1}) \in token\_out_r(q_i)$ is \emph{directly preceded} by entry $(r'_i,q_{i-1}) \in token\_in_r(q_i)$.
\end{observation}

Next, we show for $0 \leq i \leq min(k,j)$, $p_i = q_i$.

If $r_0 \neq r'_0$, without loss of generality, let $(r_0,p_0) \prec (r'_0,q_0)$, then the constraint of H2 on the order of the entries implies that there must exist an entry $en^{in} \in token\_in_r(v)$ such that $(r_0,p_0) \prec en^{in} \prec (r'_0,q_0)$. It is a contradiction to \autoref{obs:first}. Therefore, $r_0 = r'_0$. As a result, $(r_0,p_0) = (r'_0,q_0) \in token\_out_r(v)$, otherwise it is a contradiction to H1 (two entries in $token\_out_r(v)$ are not allowed to have the same round number unless they are the same entry).\\
Induction base: $v=v$ and $(r_0,p_0) = (r'_0,q_0) \in token\_out_r(v)$\\
Induction step: for $0 \leq i < min(k,j)$, if $p_{i-1}=q_{i-1}$ and $(r_i,p_i) = (r'_i,q_i) \in token\_out_r(p_{i-1})$, then $p_i = q_i$ and $(r_{i+1}, p_{i+1}) = (r'_{i+1}, q_{i+1}) \in token\_out_r(p_i)$.

The induction hypothesis implies $r_i = r'_i$ and $p_i = q_i$. Therefore, (1) both $(r_{i+1}, p_{i+1})$ and $(r'_{i+1}, q_{i+1})$ are in the same history log $token\_out_r(p_i)$ and (2) $(r_i, p_{i-1}) = (r'_i, q_{i-1}) \in token\_in_r(p_i)$. Based on \autoref{obs:second}, we have a subsequent observation: 
\begin{observation}
\label{obs:third}
both $(r_{i+1}, p_{i+1})$ and $(r'_{i+1}, q_{i+1})$ are directly preceded by the same entry $(r_i, p_{i-1}) \in token\_in_r(p_i)$
\end{observation}
If $r_{i+1} \neq r'_{i+1}$, without loss of generality, let $(r_{i+1}, p_{i+1}) \prec (r'_{i+1}, q_{i+1})$, then the constraint of H2 on the order of the entries implies that there must exist an entry $en^{in} \in token\_in_r(p_{i+1})$ such that $(r_{i+1},p_{i+1}) \prec en^{in} \prec (r'_{i+1},q_{i+1})$. It is a contradiction to \autoref{obs:third}. Therefore, $r_{i+1} = r'_{i+1}$ and $(r_{i+1}, p_{i+1}) = (r'_{i+1}, q_{i+1})$ (otherwise it is a contradiction to H1).

By the induction above, if $k = j$, then $u = w$. It is a contradiction since $u$ and $w$ are different nodes. If $k \neq j$, without loss of generality, let $k > j$, then we have $(r_{j}, p_j) = (r'_{j}, w)$ and $p_{j} = w$. Note that $(r'_{j}, w)$ is the last entry of $token\_in_r(w)$ and $s_r(w) = 1$. Therefore, there does not exist an entry $en^{out} \in token\_out_r(w)$ such that $(r_{j}, p_j) \prec en^{out}$ (by H2.2 and H2.3). It is a contradiction because $(r_{j+1}, p_{j+1}) \in token\_out(w)$ and $(r_{j}, p_j) \prec (r_{j+1}, p_{j+1})$.
\end{sketchproof}
\end{hide}

\section{Concluding Remarks}
\label{sec:discussion}

In the introduction, we mentioned quite a few implications of PLSs, the study of which for the notion of RPLS may yield interesting results. Let us consider yet an additional implication of the new notion. Under the PLS generalization studied here, the predicate was changed together with (legitimate) changes in the configuration. Still, the examples addressed only the case that the legality predicate is defined for a single configuration at a time. One could define predicates involving more than a single configuration (intuitively, this makes it easier to check also {\em liveness} properties
while checking only one configuration at a time is better suited for checking  {\em safety} ). For example, given a universal scheme that saves the whole reachable history and users inputs at the nodes (constructing a universal RPLS based on the approach of \cite{afek2002local}), one can verify that every node received the token in the last $t$ rounds for some $t$ (unless the history in all the nodes is fake, note, though, that the history in each node is eventually updated to be correct). 

Another generalization addresses the ``locality'' of the marker, and especially that of the initial configuration. 
Informally, Linial \cite{linial1992locality} asked ``from which distance must the information arrive to compute a given function''. One could ask similar questions for the checking, rather than for computing. In particular,
using the approach of \cite{afek2002local}, the labels of some node $v$ are influenced even by an event that happened at some large distance $t$ from a node $v$. It may take at least $t$ time after the event for the label at $v$ to be impacted. Let a $t$-semi-universal RPLS be one where a node maintains all the history of all the nodes (as in \cite{afek2002local}) but only the history of the last $t$ (rather than $n$) rounds. It would be interesting to characterize the hierarchy of (reactive/interactive) distributed tasks. The {\em distributed task class} Check-Local $t$ would then consist of all distributed tasks with legality predicates that can be checked by a $t$-semi-universal RPLS but not by a $(t-1)$-semi-universal RPLS. 

\vspace{-0.4cm}

\bibliographystyle{splncs04}
\bibliography{reference}
\end{document}